\documentclass{elsart}
\usepackage{psfrag}
\usepackage{exscale}                  
\usepackage[intlimits]{amsmath}       
\usepackage{amsfonts}
\usepackage{amssymb,amscd}
\usepackage[dvips]{epsfig}
\usepackage{array}
\usepackage{wrapfig}
\usepackage{graphics}
\usepackage{graphicx}
\usepackage{dcolumn}

\newcommand{\ar}{\arrowvert}
\newcommand{\ra}{\rangle}
\newcommand{\la}{\langle}
\newcommand{\da}{\dagger}

\newcommand{\be}{\begin{equation}}
\newcommand{\ee}{\end{equation}}
\newcommand{\ba}{\begin{eqnarray}}
\newcommand{\ea}{\end{eqnarray}}

\begin{document}

\begin{frontmatter}

\title{
When hadrons become unstable: a novel type of non-analyticity 
in chiral extrapolations
}
\author{
F.-K.~Guo$^1$, C. Hanhart$^2$, F. J. Llanes-Estrada$^3$, U.-G.
Mei{\ss}ner$^{1,2}$}
\address{$^1$ Helmholtz-Institut f\"ur Strahlen- und Kernphysik and Bethe Center for
Theoretical Physics, Universit\"at Bonn, D-53115 Bonn, Germany\\
$^2$ Institut f\"ur Kernphysik, Institute for Advanced Simulation and
J\"ulich Center for Hadron Physics, Forschungszentrum J\"ulich, D-52425 J\"ulich,
Germany\\
$^3$ Dept. F\'{\i}sica Te\'orica I,  Universidad
Complutense, 28040 Madrid, Spain}

%
\date{today}

\begin{abstract}
Hadron masses show a specific dependence on the quark masses. Therefore,
the variation of these masses can cause a resonance in a hadronic scattering amplitude to
become a bound state. Consequently, the amplitude exhibits a non-analytic
behavior at this transition.
Crossed amplitudes, where the resonance can be exchanged in the $t$-channel, can
be shown to exhibit the same phenomenon by $s\to t$ analytic continuation.
This entails possible kinks in lattice quark mass extrapolations needed to
compute hadronic observables.

\end{abstract}

\begin{keyword} Analyticity of S-matrix, Lattice QCD,
chiral extrapolations
\PACS 11.55.Bq  12.38.Gc
\end{keyword}

\end{frontmatter}
\section{Introduction}

It is of current interest to obtain lattice Quantum Chromodynamics (QCD)
predictions for hadronic observables, both to test QCD in the strong-coupling
regime, and to compute QCD backgrounds to new physics searches. It is customary
in these lattice gauge theory computations, due to the large numerical costs, to
perform simulations with unphysically large masses of the light quarks. Then a
smooth extrapolation formula to physical values, inspired by chiral perturbation
theory is usually employed to obtain the physical results (for a recent
review see Ref.~\cite{Bernard:2007zu}).\footnote{We are
  well aware that first simulations with physical quark masses or even less
\cite{Aoki:2009ix,Durr:2010vn} become available, but these are still exceptions.}
However, there are no theorems of S-matrix theory guaranteeing the analyticity
of such an extrapolation for larger quark masses, denoted as $m_q$ in what follows,
beyond the regime where chiral
perturbation theory is applicable.\footnote{It is well-known that certain
  non-analyticities in the quark masses can be shown to hold for arbitrary
  momenta (see Ref.~\cite{Gasser:1979hf} and references therein), but these
 are not the effects we are dealing with.}
Note that one can equally well talk of the pion mass instead of the quark mass, since
they are related by the Gell-Mann-Oakes-Renner
relation~\cite{GellMann:1968rz} $m_\pi^2 f_\pi^2 = 2m_q \la \bar q q \ra +
{\mathcal O}(m_q^2)$, where the corrections $\sim m_q^2$ are known to be very small.
On the contrary, in this letter we expose non-analyticities (kinks in the
$m_q$-extrapolation) that may arise when a resonance becomes bound upon varying
$m_q$.
For example, in the pion form factor or the $B\to \pi \pi$ decay amplitude, the
relevant resonance is the $\rho(770)$. For the weak $K\to \pi$ transition form
factor, the $K^*$ matters. Let us stress that the value of $s$ is of no concern,
the non-analyticity in the variable $m_q$ if present, affects the entire
amplitude or form factor. We will exemplify this with both time-like and space-like
pion form factors.
So far the analysis has been carried out for the
$m_q$ dependence of the resonance mass itself,
$m_\rho(m_q)$ for example~\cite{Hanhart:2008mx,Nebreda:2010if}.  Here, we
demonstrate the generality of the phenomenon affecting the computation of most
hadron observables, which completes the preliminary results reported
in~\cite{PoS}.

Our results are relevant because 
there is much active lattice research in form factor
determinations, see e.g.~\cite{Boyle:2008yd,Frezzotti:2008dr,Aoki:2009qn} and
spectroscopy~\cite{Durr:2008zz,Lin:2008pr,Baron:2010bv,Engel:2010my}.
Extrapolation formulae are available for both form factors \cite{Bijnens:1998fm}
and spectroscopy, e.g.~\cite{Bruns:2004tj}. Typically these extrapolations are smooth
except for the usual chiral logarithms of the pion mass,
$\log\left(m_\pi^2/\mu^2\right)$, that present a non-analiticity at $m_\pi=0$.

\section{Illustration: a simple model}

\begin{figure}[htbp]
\centerline{\includegraphics[width=7cm]{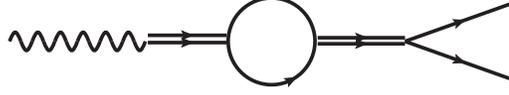}}
\caption{$\rho$-meson contribution to pion time-like form factor with one-loop
vacuum polarization. Solid, double and wiggly lines denote pions,
$\rho$-mesons and  photons, respectively.
\label{fig:diag}}
\end{figure}

To expose the feature in the simplest possible physical manner, we now
focus on amplitudes with two pions, and the role of the $\rho$-resonance. In
the next section it will be shown, however, that the results are general. 
In a simple field theory where the two pions are coupled to the resonance 
and the latter is represented as
an additional field, the threshold effect appears through the vacuum
polarization of the resonance, whose imaginary part controls the decay width.
Therefore, the size of the possible non-analyticities in any amplitude has to be
proportional to the width of the resonance and appear only in, at least,
one-loop calculations, such as depicted in Fig.~\ref{fig:diag} for the
time-like pion form factor.

We therefore proceed to study the $\pi$-$\rho$ case at one loop. The bare $\rho$-meson
propagator is $1/ (s-m_{0\rho}^2)$ --- the one-loop vacuum
polarization will renormalize the bare mass $m_{0\rho}$ to $m_\rho$.
The $s$-channel on-shell unitarization for the scattering amplitude reads
\be%
\label{amplitude} T^{11}(s) = \frac{V^{11}(s)}{1-G(s)V^{11}(s)}= \frac{
-\frac{4}{3}g_{\rho\pi\pi}^2 \ar {\bf p}\ar^2}
{z^\rho_0(s-m_{0\rho}^2)+\frac{4}{3}g_{\rho\pi\pi}^2 \ar {\bf p}\ar^2G(s)}
\ee%
in terms of the Born amplitude for $\rho\to\pi\pi$ (with $J=I=1$),
\be%
V^{11}(s)
= -\frac{4}{3} \frac{g_{\rho\pi\pi}^2}{z^\rho_0} \frac{ \ar {\bf p}\ar^2}
{s-m_{0\rho}^2} \ ,
\ee%
where the $\rho$ wave function renormalization constant $z^\rho_0$ comes from
the relation between the bare coupling constant and the renormalized one. The
factor $\ar {\bf p}\ar^2=s/4-m_\pi^2$ stems from the on-shell p-wave derivative
coupling.
With this on-shell factorization, the denominator in Eq.~(\ref{amplitude})
contains the unregularized scalar one-loop function
\be%
\label{gdef}
G(s)= \frac{1}{16\pi^2} \left(
R + \log\left( \frac{m_\pi^2}{\mu^2}\right)+1 -\bar{J}(s)
\right) \ .
\ee%
We use the convention
\be\label{jotita}%
\bar{J}(s)= 2 + \sigma \log \left( \frac{\sigma-1}{\sigma+1}
\right)
\ee%
with $\sigma=2|{\bf p}|/\sqrt{s}=\sqrt{1-{4m_\pi^2}/{s}}$ for the relativistic
phase space.
The divergence in dimensional regularization appears in
$$
R=\frac{2}{d-4}-\log(4\pi)-\Gamma'(1)-1 \ .
$$
with the number of space-time dimensions $d\to 4$.

Returning to our main issue, consider
the imaginary part of the vacuum polarization in the
denominator of Eq.~(\ref{amplitude}). It is given by
\be {\rm Im}~\Pi =
\frac{\pi\sigma}{16\pi^2}\frac{4}{3}g_{\rho\pi\pi}^2 \ar {\bf
p}\ar^2\theta(s-4m_\pi^2)
\ee
and leads to the well known non-analyticity in $s$
(branch point at $s=4m_\pi^2$).  A trivial observation is that, reciprocally,
there is a non-analyticity in $m_\pi$ for fixed $s$ at $m_\pi=\sqrt{s/4}$.
More subtle is to notice that when $2m_\pi= m_\rho(m_\pi)$ (in
App.~2, we show generally that such a situation will occur for the
$\rho$), a similar non-analyticity affects the amplitude for all $s$ due to the
dependence of the amplitude on the renormalized $m_\rho$ (physical pole
position) that suddenly changes from a bound state on the real axis to an
unbound resonance.

\begin{figure}[t]
\centerline{\includegraphics[width=8cm]{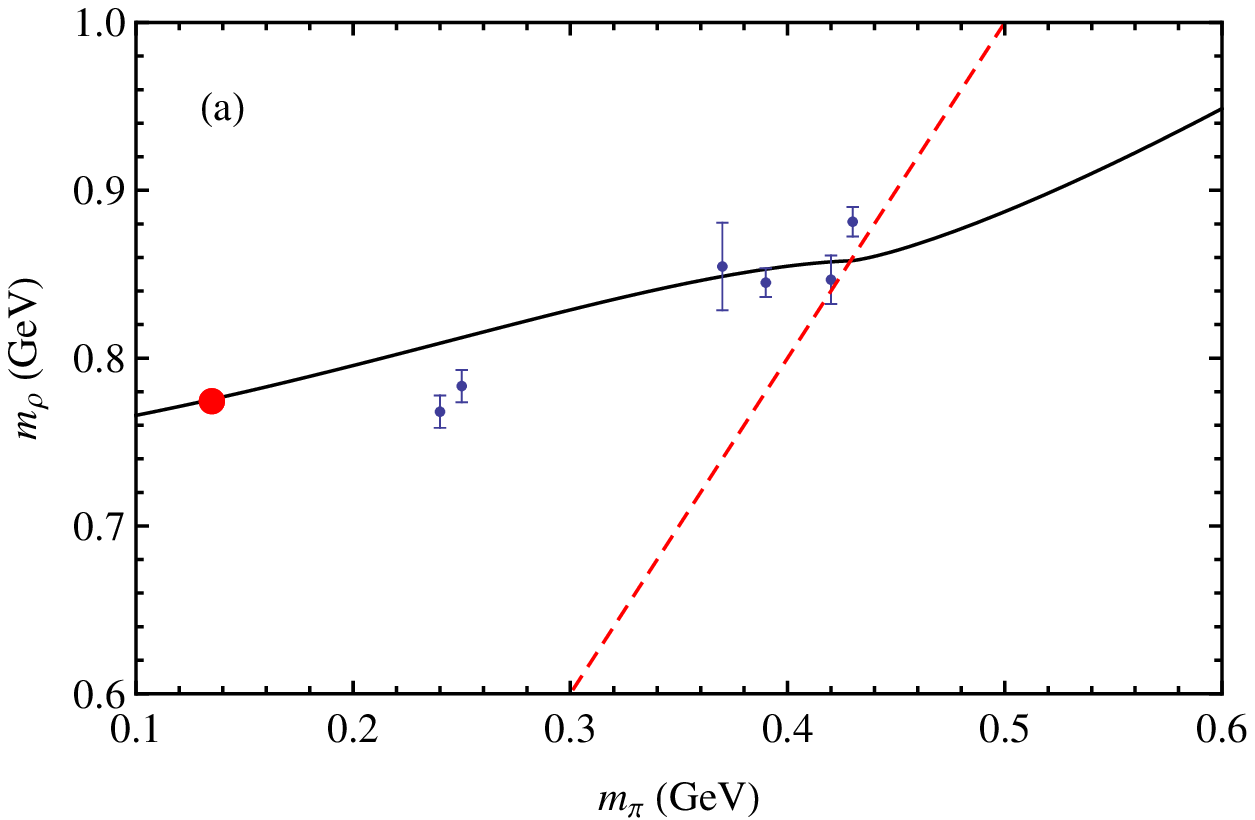}}
\vspace{2mm}
\centerline{\includegraphics[width=8cm]{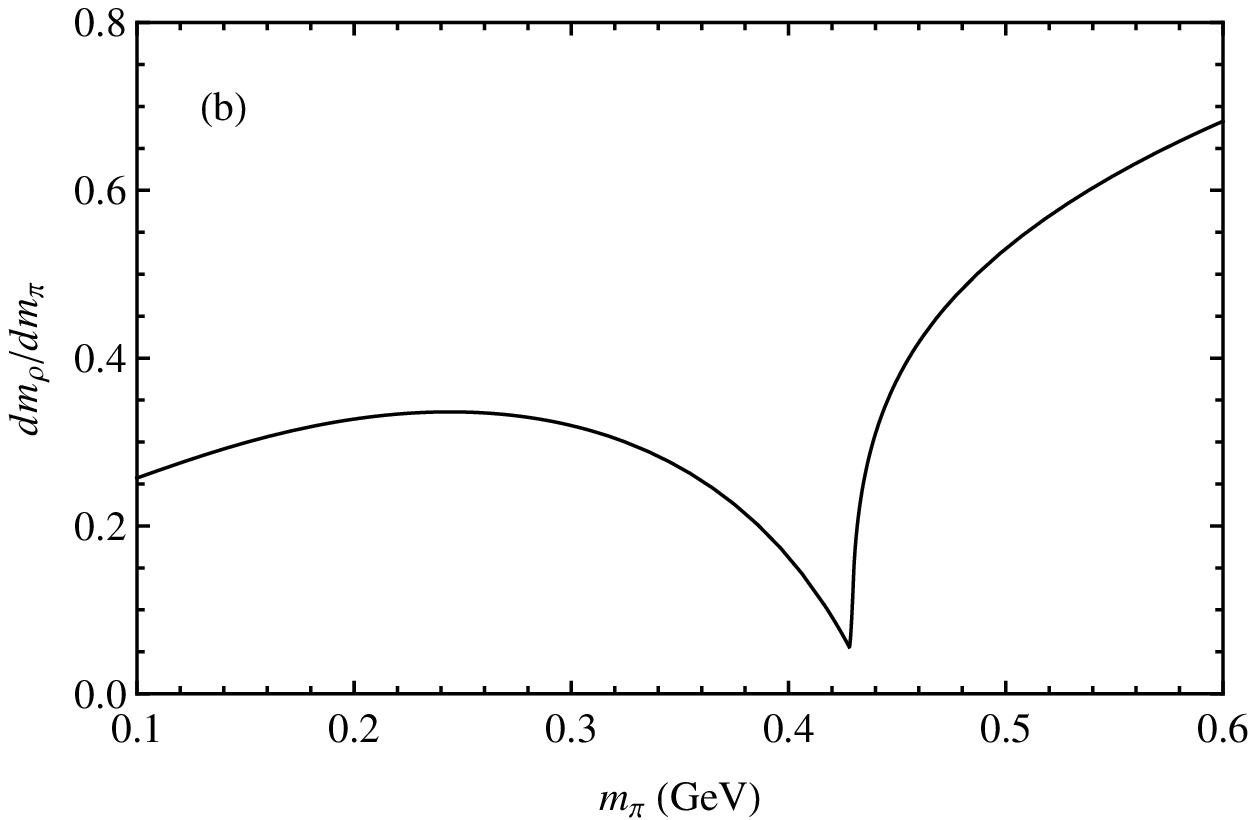}}
\caption{
Dependence of the rho mass $m_\rho$ (a) and its derivative  ${dm_\rho}/{dm_\pi}$
(b) on the pion mass obtained from Eq.~(\ref{polemass}).
The dashed line in (a) denotes the motion of the $\pi\pi$ threshold. Data  are
from a lattice calculation by the QCDSF Collaboration~\cite{Gockeler:2008kc}, and the
physical $\rho$ mass is represented by a thick circle.
\label{fig:polemass}}
\end{figure}

We present in Fig.~\ref{fig:polemass} the behavior of $m_\rho(m_\pi)$. The
results agree qualitatively with those from the more sophisticated treatment
in, e.g., Refs.~\cite{Hanhart:2008mx,Nebreda:2010if}.  Further details
of  the computation within the simple model are given in the App.~1.
Eq.~(\ref{polemass}) contains the combination
$$
\frac1{z^\rho_0} {\rm Re\ }\bar J(m_\rho^2) \left(\frac{m_\rho^2}{4}-m_\pi^2\right)
$$
that is non-analytic for the value of the pion mass where the $\rho$ becomes
bound --- c.f.
Eqs.~(\ref{jotita},\ref{zetita}). This is seen especially as a  kink in
the derivative ${d m_\rho}/{d m_\pi}$, shown in Fig.~\ref{fig:polemass}(b). Therefore, the position of the pole
in the pion scattering amplitude appears not to be an analytic function of the
pion mass.
The same phenomenon
will appear in other amplitudes, such as the time-like electromagnetic
form-factor. This can then be analytically continued (in $s$) to the
space-like  side and the same phenomenon will appear for, say, the
radius squared (which appears naturally in the low-energy expansion
of the space-like form-factor).
To illustrate this effect we now use the simplest realization of vector
meson dominance (VMD), where the direct coupling of the photon to the pion is neglected and the form
factor is entirely given by the photon-$\rho$-meson coupling to
the intermediate resonance (with the strength  $g_{\rho\gamma}$)~\cite{sakurai}.
The tree level formula
\be%
F^{\rm tree}(s) = \frac{m_\rho^2}{m_\rho^2 - s} 
\ee%
already suggests a kink if one substitutes the dependence of the pole mass
$m_\rho(m_\pi)$ that we have argued to be non-analytic. This lack of analyticity
appears then in the squared charge radius in the Breit frame $\la r^2 \ra =
6/{m_\rho^2}$.
Although the mentioned VMD description is too simplistic to exhibit all pertinent features
of the pion vector form factor, it is very useful for illustrative purposes.
In Appendix 1 it is demonstrated that the full one loop amplitude exhibits the
same features and in the next section we demonstrate that our findings are
indeed model independent.

\begin{figure}[t]
\centerline{\includegraphics[width=8cm]{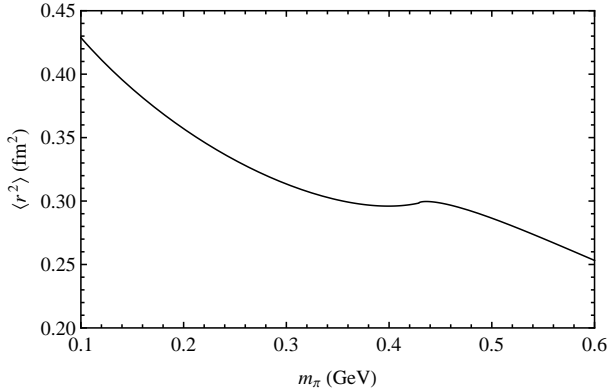}}
\caption{The squared pion charge radius also presents a kink in
its first derivative as a function of $m_\pi$, at the point where the
$\rho$ resonance becomes bound.
\label{fig:rsq}}
\end{figure}

The one-loop result for the charge radius squared as a function of the pion mass
is displayed in Fig.~\ref{fig:rsq}. We have assumed that $g_{\rho\pi\pi}$ is
independent of the pion mass.\footnote{The pion mass dependence of
$g_{\rho\pi\pi}$ is very moderate from both the unitarized chiral perturbation
theory~\cite{Hanhart:2008mx} and very recent lattice
simulations~\cite{Feng:2010es}.} 
Thus, the threshold non-analyticity when the $\rho$--resonance becomes bound is
inherited by the space-like form factor. This result may not seem intuitive,
since the argument $t$ of the space-like form factor is apparently very far from
any thresholds associated with $s$, so perhaps it is helpful to think of the
form factor as a function of two variables $F(s,m_q)$. The non-analyticity
enters because of the implicit pion--mass dependence through $m_\rho(m_q)$ and
is not affected by the analytic continuation in the other variable $s\to t$.

\section{Generalisation of the results}

We now turn to a model-independent discussion of the effect,
introduced so far within a particular model,
for a fixed $\pi\pi$ partial wave.
In a model-independent way
this effect can be studied by employing an Omn\`es representation for the
form-factor as given e.g. in
Refs.~\cite{Gasser:1990bv,yndurain,Oller:2007xd,Ananthanarayan:2011xt}.
This renowned relation expresses
the form factor as an
integral over the scattering phase shift. In once-subtracted form it reads in
the absence of bound states
\be\label{omnesrep2} %
F(t,m_\pi^2) =\Omega(t,m_\pi^2) = \exp\left(\frac{t}{\pi} \int_{4m_\pi^2}^\infty ds
\frac{\delta_{11}(s,m_\pi^2)}{s(s-t-i\epsilon)} \right) \ .
\ee%
Then the charge radius in the Omn\`es representation is expressed in terms of  the
$\pi\pi$ scattering phase shift as~\cite{Oller:2007xd}
\be \label{radiusfromshift} %
\la r^2 \ra = \frac{6}{\pi} \int_{4m_\pi^2}^\infty ds
\frac{\delta_{11}(s,m_\pi^2)} {s^2} \ .
\ee%
In the presence of a bound state there is an additional singularity on the
first sheet and thus the dispersion integral needs to be modified. It
now reads~\cite{Barton},
\be%
F_b(t,m_\pi^2)=\left(1+\frac{t g_{\gamma\rho}g_{\rho\pi\pi} }{m_\rho^2(m_\rho^2-t)}
\frac{1}{\Omega(m_\rho^2,m_\pi^2)} \right) \Omega(t,m_\pi^2) \ .
\ee%
Eq.~(\ref{radiusfromshift}) needs to be adapted accordingly
\be \label{radiusfromshift_mod} %
\la r^2 \ra_b = \frac{6}{\Omega(m_\rho^2,m_\pi^2)}\frac{g_{\gamma\rho}g_{\rho\pi\pi} }{m_\rho^4}
+\frac{6}{\pi} \int_{4m_\pi^2}^\infty ds
\frac{\delta_{11}(s,m_\pi^2)} {s^2} \ .
\ee%
Here, we introduced the subscript $b$ to distinguish the quantities defined in
the presence of a bound state from those given in Eqs.~(\ref{omnesrep2},\ref{radiusfromshift}).
\begin{figure}[t]
\centerline{\includegraphics[width=8cm]{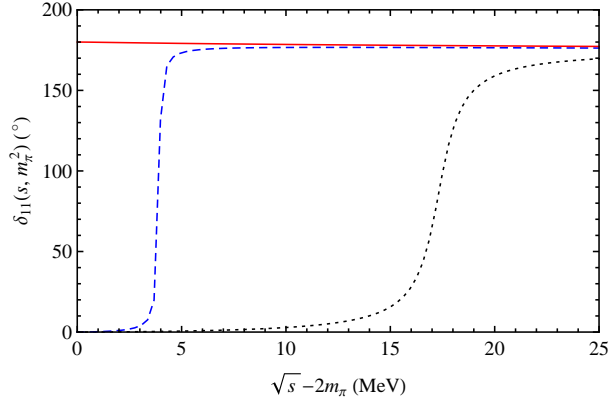}}
\caption{Pion mass dependence of $\delta_{11}(s,m_\pi^2)$ from the
  one--loop model, as 
the pion mass approaches the value where the $\rho$ becomes stable. Shown are
the phases for $m_\pi=$420~MeV (dotted line), 427~MeV (dashed line) and
431~MeV (solid line).
\label{fig:critphases} }
\end{figure}
It should be stressed that the form factor is continuous at the value of the
pion mass, where the $\rho$ becomes a stable state, for all $t\neq m_\rho^2$. 
To see this we first observe
that the integrals over the phases in 
Eqs.~(\ref{radiusfromshift},\ref{radiusfromshift_mod}) converge towards 
each other, as the $\rho$--mass
approaches the two--pion threshold. This follows directly from the behaviour
of the phases shown in Fig.~\ref{fig:critphases} --- as soon as the
$\rho$ appears as a stable state at $m_\pi=430\,$MeV, according to Levinson's
theorem the phase shift starts from $\pi$. In addition, when approaching
 the point 
 $m_\rho(m_\pi)=2m_\pi$ from larger pion masses one finds
\ba%
\log \Omega(t,m_\pi^2) \simeq \frac{t}{\pi}\delta_{11} \int_{4m_\pi^2}^\infty ds
\frac1{s(s-t-i\epsilon)} = \frac{\delta_{11}}{\pi} \log \frac{4 m_\pi^2}{|4
m_\pi^2-t|} ,
\ea%
where we used that in this limit $\delta_{11}$ is a slowly varying function
of
$s$ in the energy range of interest.
Evidently, $1/\Omega(t,m_\pi^2)$ vanishes, when $t=4m_\pi^2$. 
However, as becomes apparent in Eq.~(\ref{radiusfromshift_mod}),
the non--analyticity in, e.g., $m_\rho$ directly influences the quark
mass dependence of the squared radius~\cite{Guo:2008nc}.

\section{Summary and conclusions}

In this work we discussed a non--analyticity in the chiral extrapolation of
physical quantities that emerges when due to a change in the quark mass a
state transforms from a resonance --- poles on the second sheet --- to a
physical state with a pole on the first sheet. We established that the
analytic continuation in the kinematic variable $s\to t$
does carry over the non-analyticity in $m_\pi$ from the time-like to the
space-like domain. This kind of behavior is model-independent.

We have analyzed in this letter the case of the pion form factor, but the  same
phenomenon should appear in other form factors. For example, in the $K\to\pi$
weak vector transition form factor \cite{Antonio:2006ev,Becirevic:2004ya}, when
$m_\pi\simeq 350$~MeV the $K^*$ resonance should also become bound, and develop
a non-analyticity.

The situation is even more interesting for scalar form factors, where one has a
kink directly in the function (be it the mass or the squared radius) and not in
its derivative with respect to the pion mass. This is simply  because the factor
$\ar {\bf p}\ar^2$ from the $J=1$ derivative coupling is absent (as already
discussed  in detail in Ref.~\cite{Gasser:1990bv}). Then it will be easier for
lattice data to isolate such a structure (that is not yet visible in existing
simulations, see e.g.~\cite{Aoki:2009qn}). If a relative drop $(\delta
F^{(n)})/F^{(n)}$ in the $n^{\rm th}$ derivative of a function is to be
identified visually, the error acceptable in the lattice computation of $F$
itself is, as a rule of thumb, $(\delta F)/F \le (\delta F^{(n)})/(2^n F^{(n)})$
since each derivative with a good mid-point numerical method requires two
evaluations. Hence, we would propose that the scalar pion form factor be
computed with smaller statistical error bars and smaller $t$-intervals, as a
favorable system to try to find the non-analyticity, given that there is no
phase-space suppression and that the coupling $g_{\sigma\pi\pi}$ is large.
Calculations using unitarized chiral perturbation theory predict that the
$\sigma$ meson becomes bound at about $m_\pi\simeq
350$~MeV~\cite{Hanhart:2008mx}.
Full QCD simulations for scalar quantities at sufficiently low
pion masses will, however, not appear in the near future for those are a lot more
computer time intensive compared to the ones discussed due to the presence of
disconnected diagrams.

We now  examine to what extent non-analyticities have been stressed  in
earlier studies. Very old work focused on the particle virtuality for fixed mass, be
it in perturbation theory or with the Lehman representation
\cite{Dremin:1961zza}, or for scaling deeply-inelastic scattering functions
\cite{Geshkenbein:1973ai}. In both cases the phenomenon of a resonance becoming
a bound state during the particle mass variation is absent, and those authors
found analyticities in the transferred momentum plane with the physical values
of the particle masses.  Closer in spirit to our work, features in quark--mass 
extrapolations due to presence of a threshold have already been discussed 
in Ref.~\cite{Bernard:2007cm}, which focuses on the avoided level crossing in
a finite volume. A small cusp in the pion mass
dependence of the mass of the $\Delta$ resonance~\cite{Bernard:2009mw} is produced.
In Ref.~\cite{mainz} it was stressed that the kind of non-analyticity discussed
in Ref.~\cite{Bernard:2009mw} also shows up in electro-magnetic properties such
as the magnetic moment. The effect we discussed introduces an {\it
additional} non-analyticity in the radii.
Finally, another  kind of non-analyticity --- possible discontinuities
--- in the chiral extrapolation of hadron masses was 
proposed in~\cite{Semke:2007zz}.

The non-analiticity that we uncover is a feature of continuum field-theory, 
and not an artifact of lattice--quantization.
It is possible that accurate lattice data should be able to isolate
these non-analyticities, provided the volume is large enough that the
resonance is not bound by the minimum momentum possible on the lattice 
(though the alternative non-analiticities of Ref.~\cite{Bernard:2009mw} mask the effect).
They should be taken into account when attempting to extrapolate lattice
data to physical pion masses when high precision is expected.

\bigskip

{\emph{ We thank Stephan D\"urr and Akaki Rusetsky for useful
discussions. This work was supported in part by grants provided by the HGF to
the virtual institute ``Spin and strong QCD'' (VH-VI-231), the DFG (SFB/TR 16)
and the EU I3HP ``Study of Strongly Interacting Matter'' under the Seventh
Framework Program of the EU, FPA 2008-00592, 2007-29115-E, FIS2008-01323 (Spain)
and 227431, HadronPhysics2 (EU). FJLE thanks the members of the Nuclear Theory
Center at Forschungszentrum J\"ulich for their hospitality during this work. UGM also thanks the BMBF for support (Grant No. 06BN9006).}}

\section*{Appendix 1: One-loop renormalization of the  $\rho\pi\pi$ model.}
\label{App1}

Our choice of renormalization is meant to expose the pole mass and decay
coupling constant in the amplitude, so this one is expressed in terms of
directly measurable quantities. To achieve this, we add and subtract to the
denominator of Eq.~(\ref{amplitude}) the vacuum polarization evaluated at the
(still unknown) $\rho$ pole mass, that is,
$$
\frac{4}{3}g_{\rho\pi\pi}^2 G(m_\rho^2)\left(\frac{m_\rho^2}{4}-m_\pi^2\right)\ .
$$
Imposing now the renormalization condition that the position of the pole
in the denominator be at $m_\rho$ yields the equation
\be \label{polemass}%
m_\rho^2= m_{0\rho }^2 - \frac{4}{3}\frac{g_{\rho\pi\pi}^2}{ z^\rho_0}
G(m_\rho^2)\left(\frac{m_\rho^2}{4}-m_\pi^2\right) \ .
\ee%
Above the two-pion threshold, $m_\rho$ is complex. But in the numerical
calculations, for simplicity, we take $m_\rho$ to be real.
This amounts to neglecting ${\rm Im}~G$ --- which is a very good
approximation near the kink where phase space closes.
Note that $G$ contains an infinity that needs to be absorbed into the
bare mass. Since the divergence is multiplied by
$(m_\rho^2/4-m_\pi^2)/z^\rho_0$ the subtraction procedure
calls for introducing a pion--mass dependent mass term.
Thus, since we want to keep the pion mass dependences explicit, the Lagrangian density for the
model needs to contain a counterterm proportional to $m_\pi^2 \rho^\da
\rho$.

We therefore define the renormalized
mass and its mass derivative with respect to $m_\pi^2$ through
\be%
\bar{m}_{0\rho }^2 + \bar{m}_{0\rho }^{'\ 2} m_\pi^2 = m_{0\rho }^2 -
\frac{4}{3}\frac{g_{\rho\pi\pi}^2}{ z^\rho_0} \left(\frac{m_\rho^2}{4}-m_\pi^2\right)
\frac{R+1}{16\pi^2}
\ee%
and demand that the two constants $\bar{m}_{0\rho }^2$ and $\bar{m}_{0\rho }^{'\
2}$ be pion-mass independent.
These two parameters can be fixed using the physical $\rho$ meson mass at the
physical point for $m_\pi$ and lattice data.
We choose as renormalization scale, which enters Eq.~(\ref{polemass})) through $G(m_\rho^2)$
(c.f. Eq.~(\ref{gdef})), $\mu={m}_{\rho}$, the pole mass itself. Then $z^\rho_0$ becomes known
(see below), and Eq.~(\ref{polemass}) can be solved. The best fit to the lattice
data from the QCDSF Collaboration~\cite{Gockeler:2008kc} with the constraint from
the physical rho mass gives $\bar{m}_{0\rho}=0.707$~GeV, and
$\bar{m}_{0\rho}^{\prime}=1.13$.



At this point we have guaranteed that the pion-pion scattering
amplitudes has a pole at physical $m_\rho$ for the physical pion mass, and
we can compute the variation of the pole position with the pion mass if
this dependence is known for $z^\rho_0$, so we also need to solve for it.

The second renormalization condition we impose is that $g_{\rho\pi\pi}$ be the
physical coupling at the $\rho$-pole, obtainable from the residue of the pion
scattering amplitude
\be {\rm Res \ } T^{11}(s) = \lim_{s\to m_\rho^2}
(s-m_\rho^2)\, T^{11}(s) \ . \ee
Imposing that the residue be
\be {\rm Res \ }
T^{11}(s) = -\frac{4}{3} g_{\rho\pi\pi}^2 \left(\frac{m_{\rho}^2}{4}-m_\pi^2\right)
\ee
and taking into account that, in terms of the pole mass, we have
\be
\label{amplitudepole} T^{11}(s) = \frac{ -\frac{4}{3}g_{\rho\pi\pi}^2 \ar {\bf
p}\ar^2} {z^\rho_0(s-m_{\rho}^2) + \Delta(s)+i{\rm Im}~\Pi(s)}\ee
with
\be
\Delta(s) \equiv {\rm Re}~\Pi(s) - {\rm Re}~\Pi(m_\rho^2)\ ,
\ee
where
$$\Pi(s)=\frac{1}{12\pi^2} g_{\rho\pi\pi}^2\bar J(s)\left(\frac{s}{4}-m_\pi^2\right).$$
We find
\be%
\label{zetita} z^\rho_0 = 1-\frac{d\Delta(s=m_\rho^2)}{ds}\ .
\ee %
Note the $m_{\rho}$ in the
last expression is not the physical value but $m_{\rho}(m_\pi)$, to guarantee
that $m_{\rho}(m_\pi)$ is always defined as the pole mass in the propagator.
As discussed below Eq.~(\ref{polemass}), although $m_\rho$ is complex, for
simplicity we take $m_\rho$ to be real.


Resumming the Dyson series originating from the vacuum polarization of the
$\rho$-meson one obtains for the form factor
\be%
F(s) = \frac{-g_{\rho\gamma}g_{\rho\pi\pi} \sqrt{z_0^A/z_0^\rho}}
{z_0^\rho(s-m_\rho^2)+ \Delta(s) +i {\rm Im}\Pi(s)}\ .
\ee%
The denominator is of course the same as in Eq.~(\ref{amplitudepole}), and since
the numerator is real this guarantees the same phase for form  factor and
scattering amplitude. In order to get the proper normalization in this
most simple formulation of VMD one needs to impose
\be%
g_{\rho\gamma}g_{\rho\pi\pi} \sqrt{z_0^A/z_0^\rho}=z_0^\rho m_\rho^2 - \Delta(0)
\ee%
on the photon-rho coupling. The resulting form factor
\be%
F(s)= \frac{-z_0^\rho m_\rho^2+ \Delta(0)} {z_0^\rho(s-m_\rho^2) + \Delta(s) +i
{\rm Im}\Pi(s)}
\ee%
satisfies now $F(0)=1$ and has the correct unitarity cut.
\begin{figure}[t]
\centerline{\includegraphics[width=8cm]{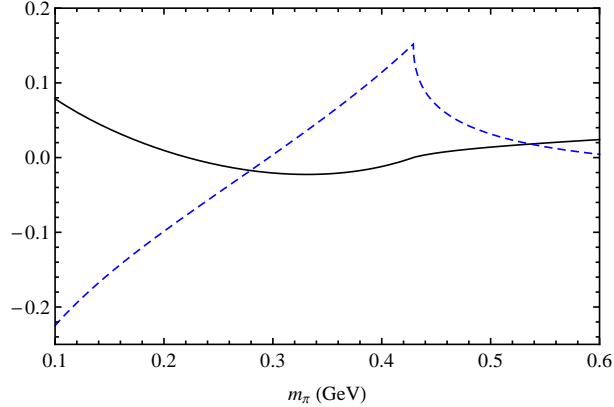}}
\caption{Pion mass dependence of auxiliary
$\Delta(0)$ (solid) and $\Delta'(m_\rho^2)=1-z_0^\rho$ (dashed). 
\label{fig:delta} }
\end{figure}
Through all the one-loop quantities $m_\rho$, $z_0^\rho$ and $\Delta$, as shown
in Fig.~\ref{fig:polemass} and Fig.~\ref{fig:delta}, the form-factor acquires a
non-analyticity in $m_\pi$. This non-analyticity appears at one loop and is
therefore proportional to $g_{\rho\pi\pi}^2$ and hence the physical resonance
width. 
Using Eq.~(\ref{zetita}) the squared charge radius becomes now
\be\label{radius}%
\la r^2 \ra = 6 \frac{z_0^\rho+ \Delta'(0)}{z_0^\rho m_\rho^2-\Delta(0)} \ .
\ee%
The derivative of $\Delta(s)$ at $s=0$ is a constant
$\Delta'(0)=-g_{\rho\pi\pi}^2/(72\pi^2)$.

\section*{Appendix 2: Position of the point $m_\rho(m_\pi)=2m_\pi$}
\label{App2}

In this appendix, we will show that the $\rho$ mass grows slower than the
two-pion threshold when increasing the pion mass, and hence there must be a
certain point after which the $\rho$ will be below the two-pion threshold.

Expanding the $\rho$ mass in terms of $m_\pi$, to the order ${\mathcal O}(p^2)$ (it
is sufficient for our purpose to work to this order; for the expansion to higher orders,
see~\cite{Bruns:2004tj}), one has
\be%
\label{aeq:mrho}
m_\rho(m_\pi) = m_{\rho0} + c_1 m_\pi^2 = m_{\rho0} + 2c_1 B_0\hat{m},
\ee%
where $m_{\rho0}$ is the rho mass in the chiral limit, $c_1$ is a low-energy
constant related to the quark mass term in the chiral expansion, $B_0=-\langle0|\bar
qq|0\rangle/f_\pi^2$ and $\hat{m}=(m_u+m_d)/2$. Generally, since the rho has
a non-vanishing (and not small) mass
even in the chiral limit, one has $m_{\rho0}>2m_\pi$ for small values of the
pion mass. The points where $m_\rho$ coincides with $2m_\pi$ are then simply
given by the solutions of $m_{\rho0} + c_1 m_\pi^2 = 2m_\pi$, i.e.
\be%
\label{aeq:meetpoint} m_\pi = \frac1{c_1}(1\pm\sqrt{1-c_1m_{\rho0}}).
\ee%
Hence a crossing happens if and only if
\be%
c_1 \leq \frac1{m_{\rho0}}\ .
\ee%
To determine $c_1$ we resort to quark-mass controlled $SU(3)$ breaking, 
and expand the mass of the $K^*$ in analogy withEq.~(\ref{aeq:mrho})
\be \label{aeq:mkstar}
m_{K^*}(m_\pi) = m_{\rho0} + c_1 B_0(m_s+\hat{m}) = m_{\rho0} + c_1 m_K^2. 
\ee%
 Because $c_1$ is independent of the quark mass by definition, 
it can be used for unphysical pion masses after determining it
using physical meson masses by
\be%
c_1 = \frac{m_{K^*}-m_\rho}{m_K^2-m_\rho^2} = 0.51~{\rm GeV}^{-1}.
\ee%
Therefore, as long as $m_{\rho0}\leq 1960$~meV, the inequality $c_1 \leq
1/m_{\rho0}$ can be fulfilled. It is believed that $m_{\rho0}$ is not far from
the physical mass $m_\rho=770\,$MeV, so that the rho mass will coincides with
$2m_\pi$ at some value(s) of $m_\pi$. One can even estimate that value. Taking,
e.g., $m_{\rho0}\approx 700$~meV, the crossing point will be at around
$m_\pi\approx 400$~meV. Finally, we note that the second solution of 
Eq.~(\ref{aeq:meetpoint}) is far beyond the applicability region of 
the chiral expansion.


\end{document}